\include{graphicsx}

\documentclass[12pt,preprint]{aastex}

\slugcomment{}

\shorttitle{}
\shortauthors{Siegler, Close, Cruz, Martin, \& Reid}

\begin{document}


\title{Discovery of Two Very Low-Mass Binaries: Final Results of an Adaptive Optics Survey of Nearby M6.0-M7.5 Stars}

\author{Nick Siegler\altaffilmark{1}, Laird M. Close\altaffilmark{1}, Kelle L. Cruz\altaffilmark{2,3}, Eduardo L. Mart\'\i n\altaffilmark{4}, \& I. Neill Reid\altaffilmark{2,5}}
%\affil{Steward Observatory, University of Arizona}
%\email{nsiegler@as.arizona.edu}

\altaffiltext{1}{Steward Observatory, University of Arizona, 933 N. Cherry Ave., Tucson, AZ 85721, USA}
\altaffiltext{2}{Department of Physics and Astronomy, University of Pennsylvania, 209 South 33rd Street, Philadelphia, PA 19104, USA}
\altaffiltext{3}{American Museum of Natural History, Department of 
Astrophysics, Central Park West at 79th St., New York, NY, 10023 USA}
\altaffiltext{4}{Instituto de Astrofisica de Canarias, La Laguna, Tenerife, E-38200 Spain}
\altaffiltext{5}{Space Telescope Science Institute, 3700 San Martin Drive, Baltimore, MD 21218, USA}

\begin{abstract}
We present updated results of a high-resolution, magnitude limited
(K$_s<12$\,mag) imaging survey of nearby low-mass M6.0-M7.5 field
stars. The observations were carried out using adaptive optics at the
Gemini North, VLT, Keck II, and Subaru telescopes. Our sample of 36
stars consists predominantly of nearby ($\lesssim30$\,pc) field stars,
5 of which we have resolved as binaries. Two of the binary systems,
2MASSI J0429184-312356 and 2MASSI J1847034+552243, are presented here
for the first time. All 5 discovered binary systems have separations
between 0$\farcs08-0\farcs53$ (2\,-\,9\,AU) with similar mass ratios
(q\,$>$\,0.8, $\Delta$K$_s<1$\,mag). This result supports the
hypothesis that wide (a\,$>20$\,AU) very low-mass
(M$_{tot}\,<\,0.19\,$M$_\sun$) binary systems are rare. The projected
semimajor axis distribution of these systems peak at $\sim5$\,AU and
we report a sensitivity-corrected binary fraction of 9$_{-3}^{+4}\%$
for stars with primaries of spectral type M6.0-M7.5  with separations
$\ga3$\,AU and mass ratios q\,$\ga0.6$. Within these instrumental
sensitivities, these results support the overall trend that both the semimajor axis distribution and binary fraction are a function of the mass of the primary star and decrease with decreasing primary mass. These observations provide important constraints for low-mass binary star formation theories.

\end{abstract}

\keywords{instrumentation:\,adaptive optics---binaries:\,general---stars:\,low mass---stars:\,individual ()}

\section{Introduction}
One of the main motivations for measuring the binary fraction of stars
is to better understand the process of star formation itself. After
all, stars like our own Sun are preferentially produced in multiple
systems \citep{duq91}. The classic stellar formation mechanism of
molecular cloud core collapse and fragmentation, however, has a hard
time explaining the tightest systems. While this mechanism can explain
wide binary systems (semimajor axis $\gtrsim10$\,AU), it has some
difficulties explaining the formation of tight systems
\citep{bat02}. Additionally, the multiplicities of the lowest mass
stars and brown dwarfs appear to be statistically different from those
of more massive systems (\cite{clo04} and references within). These
differences, if proven to be real, provide important clues and
constraints to theoretical stellar formation models. 

The continuously improving statistics of binary stars brings clarity to the paradigm that the binary fraction and semimajor axis distribution are functions of the central star mass. Surveys of G dwarfs estimate a multiplicity fraction of approximately 50$\%$ for separations of 3\,AU and greater \citep{duq91}. Other surveys of similar sensitivity to systems wider than 3\,AU have found that early M dwarfs (M0-M4) have measured binary fractions of $\sim32\%$ \citep{fis92} while late M/early L dwarfs (M8.0-L0.5) estimate fractions $\sim15\%$ \citep{clo04}. The trend appears to continue to the coolest objects - L dwarfs reporting $\sim10$\,-\,15$\%$ \citep{bou03,giz03} and T dwarfs at $\sim10\%$ \citep{ber03}. The same surveys infer semimajor-axis separations to also be a function of primary mass. While G and early M dwarfs (M0-M4) show broad separation peaks of $\sim30$\,AU, late M ($\geq$\,M8), L, and T dwarfs appear to have separations tightly peaked at $\sim4$\,AU \citep{clo04}. A similar result has been shown to apply to the sequence of members in the Pleiades cluster, from solar-type stars to brown dwarfs \citep{mar03}. Together, these results are providing both clues and empirical constraints on star formation mechanisms as well as potentially help calibrate the mass-age-luminosity relation for very low-mass (VLM) stars.

In \cite{sie03}, hereafter referred to as Paper I, we presented
results from the largest flux limited (K$_s<12$\,mag) survey of nearby
field M6.0-M7.5 dwarfs. The binary fraction of this narrowly defined
spectral type range had not been quantified as those of stars slightly earlier
\citep[M0-M4;][]{fis92} and later
\citep[M8-L0.5;][]{clo04}. Considering the differences in binary
charactersitics as a function of mass as discussed above, would M6.0-M7.5 binaries have intermediary characteristics to their main
sequence neighbors or would they resemble the ultracool dwarfs?

Paper I's sample consisted of 30 stars and presented the discovery of three new binary systems using the University of Hawaii visitor AO system Hokupa'a \citep{gra98} at the Gemini North telescope. The discoveries followed characteristics of other VLM binary systems, namely relatively equal mass components (q$\,>$\,0.8) with projected separations less than 16\,AU. In this paper we present our two latest binary discoveries from this spectral range, 2MASSI J0429184-312356 and 2MASSI J1847034+552243, hereafter referred to as 2M\,0429 and 2M\,1847. These binaries were discovered with the VLT and the Subaru AO facilities, respectively. The total M6.0-M7.5 sample size is increased to 36 and we update the binary fraction results with those presented in Paper I. We present our observations and results in the following section and examine the systems' derived characteristics such as distance, age, temperature, spectral type, and mass in \S 3. In \S 4 we conclude by discussing the binary frequency and separation distribution of M6.0-M7.5 dwarfs.

\section{Observations and Results}
\subsection{The Sample}\label{bozomath_a}
We selected a flux-limited sample of 36 objects consisting of M6.0-M7.5 dwarfs with K$_s<12$\,mag and J-K$_s>0.95$\,mag from mainly 2MASS stars listed in \cite{cru03}, \cite{rei02}, and \cite{giz00}. Paper I discusses the first 30 observations and we report here the most recent 6. One of the 6 targets is a recently discovered high proper motion M dwarf - SO 025300.5+165258 \citep[$\sim3.6$\,pc away;][]{tee03}. We discuss this star further in \S 2.5. We also note that we observed at Subaru on 2003 July 10 (UT) another recently-discovered high proper motion M dwarf 2MASSI J1835379+325954 \citep[$\sim5.7$\,pc away;][]{rei03}. While not part of this sample due to its later spectral type (M8), it was observed  at high resolution and found to have no q\,$>0.8$ companions at separations $>0\farcs1$.

\subsection{The Telescopes and their AO Systems}\label{bozomath_b}
The 30 targets from Paper I were all observed at the Gemini North
telescope. The 6 targets presented here were observed at the Subaru
and VLT Observatories. Due to its recent discovery and proximity, we
conducted additional long integrations of SO 025300.5+165258 with the
Keck II telescope. Interestingly, the AO systems at these telescopes
represent the three major wavefront sensor (WFS) technologies
currently in use today. Gemini North, at the time of our observations,
and Subaru, currently, use 36-element curvature WFSs, the VLT has an infra-red Shack-Hartman, and the Keck II utilizes a visible Shack-Hartman. This survey provided the opportunity to compare and contrast how different AO WFSs differ in their abilities to lock on faint targets.

As discussed in detail in Paper I, one of the challenges in utilizing AO is locating sufficiently bright guide stars near enough one's science objects to minimize  uncertainty in the image quality introduced by isoplanicity. This is best achieved when using the target object itself as the AO guide star. This shifts the criterion of target selection from the availability of bright natural guide stars (R$\lesssim$15\,mag) to the sensitivity of the respective telescope's AO system, in particular the WFS. This becomes quite important because the probability of locating a R\,=\,15\,mag star within 30\arcsec of one's science target is only about $\sim15\%$ \citep[Fig. 3.10;][]{rod99}. The ability to guide on fainter stars also allows for both larger sample sizes and improved contrast ratios.

We were able to observe the faintest of our targets (V$\sim$19.0\,-\,19.5\,mag, I$\sim$15.5\,-\,16\,mag) only with the former Gemini North AO system Hokupa'a where we conducted the majority of the observations. No other current AO system can lock onto such faint targets. Hokupa'a, decommisioned in 2003, was a curvature-based AO system which employed in its WFS red-sensitive, photon-counting avalanche photodiodes with effectively zero read-noise. Consequently, this type of sensor is ideally suited for guiding on intrinsically faint objects as long as they are relatively red (V-I\,$\sim4$\,mag). The Keck II telescope with a more traditional Shack-Hartman WFS allows for improved angular resolution with higher obtainable Strehl ratios but requires brighter targets (V$\sim15$\,mag). We compare and contrast the performance of these two types of WFS technologies in \cite{sie02}. At both Subaru and the VLT we were able to lock on our faint low mass targets with I\,$\lesssim15.2$\,mag (K$_s\lesssim11.2$\,mag)

\subsection{Observations} \label{bozomath_c}
The two discovered binary systems, 2M\,0429 and 2M\,1847, were detected at the VLT and Subaru observatories on 2003 February 13 (UT) and 2003 July 10 (UT), respectively. A total of 6 dwarfs from our sample were observed during these two runs. Table \ref{tbl-1} lists the 4 low-mass dwarfs observed with no likely physical companion detections between $\sim0\farcs1$\,-\,15$\arcsec$. For completeness we also include the 27 single stars observed in Paper I. Table \ref{tbl-2} lists the observable properties of the 2 new binary systems along with the 3 systems presented in Paper I. Target stars were considered ``observed'' when a minimum corrected FWHM of $\sim0\farcs15$ in H band was achieved.

Each of the observations were made by dithering over 4 different quadrant positions on the infrared camera detector. For all targets we obtained both unsaturated H or K$_s$ images ($\leq10$\,s, ``short'' images), depending on seeing conditions, and saturated H images (30\,s, ``deep'' images) to gain sensitivity to potential faint companions. 

At Subaru we used the Coronagraphic Imager with AO (CIAO) without using the coronagraphic mask feature. The detector is a 1024$\times$1024 ALADDIN II InSb infrared hybrid array with a platescale of 0\farcs0217\,pixel$^{-1}$ \citep{tam00}. For 2M\,1847 we took a total of 12$\times$10\,s short exposures at H and K$_s$, 12$\times$20\,s at J, and 12$\times$60\,s deep exposures at H. At the VLT we observed with the Nasmyth AO System/NIR Imager and Spectrograph (NACO) system on UT4 (Yepun) which contains a 1024$\times$1024 ALADDIN II InSb infrared hybrid array detector with a platescale of 0\farcs0271\,pixel$^{-1}$. NACO is unique in that it utilizes an infrared WFS. We found that the infrared WFS was most efficient for objects with K$_s\leq$11.2\,mag. For 2M\,0429 we took a total of 16$\times$0.5\,s short exposures at H, 8$\times$0.5\,s at K$_s$, 12$\times$1\,s at J, and 12$\times$30\,s deep H frames.

%\placetable{tbl-1}
%\placetable{tbl-2}
\subsection{Reduction} \label{bozomath_d}

\begin{figure}
\includegraphics[angle=0,width=\columnwidth]{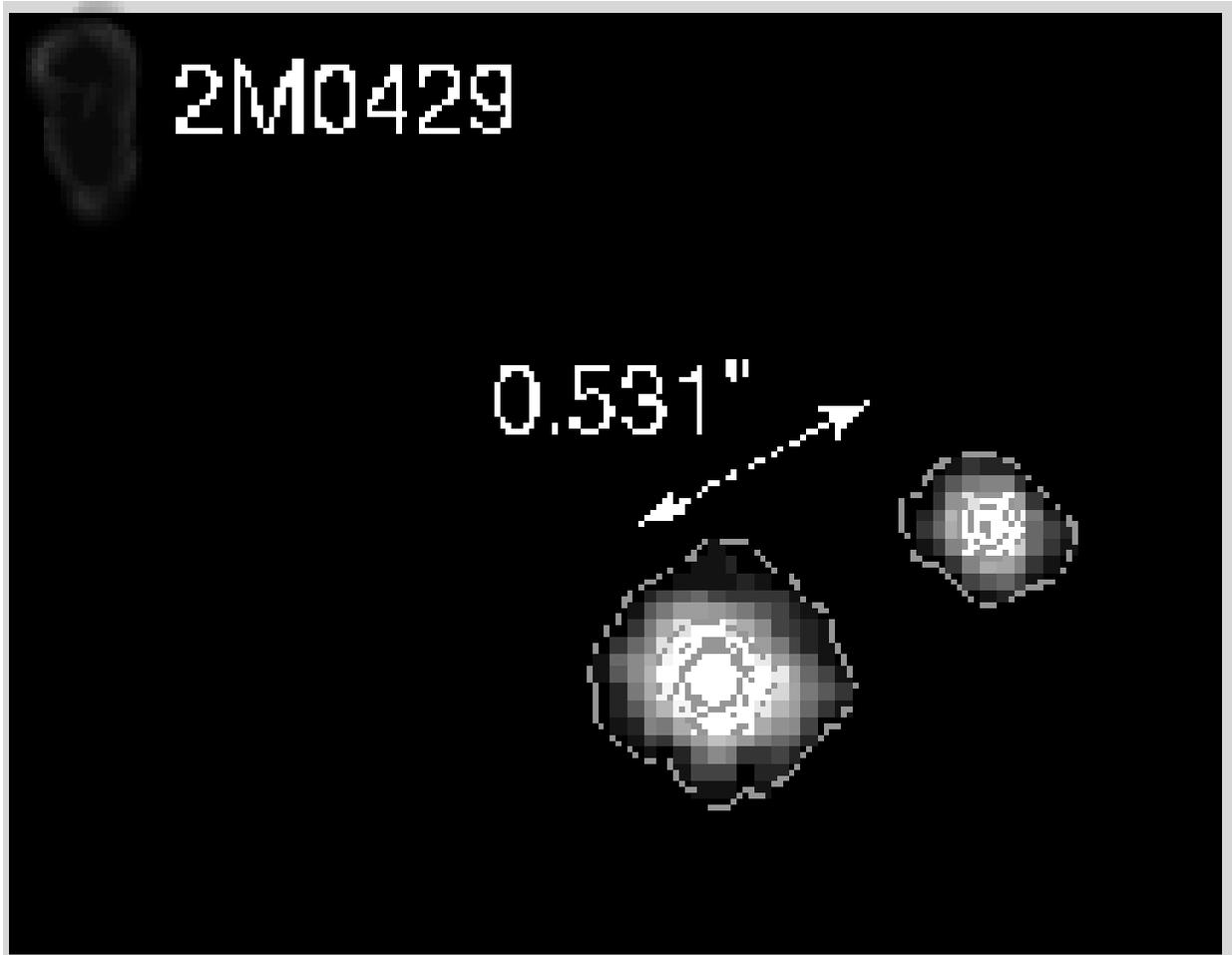}
\caption{An 8$\times$0.5\,s image of the newly discovered binary
system 2M\,0429 shown in the K$_s$ band; observed on 2003
February 13 (UT) at the VLT. The platescale is
0$\arcsec$.0271\,pixel$^{-1}$. The contours are linear at the 80, 60,
40, and 20\% levels; north is up and east is to the left.
%both_binaries.eps=f1.eps
\label{fig1}} 
\end{figure}

\begin{figure}
\includegraphics[angle=0,width=\columnwidth]{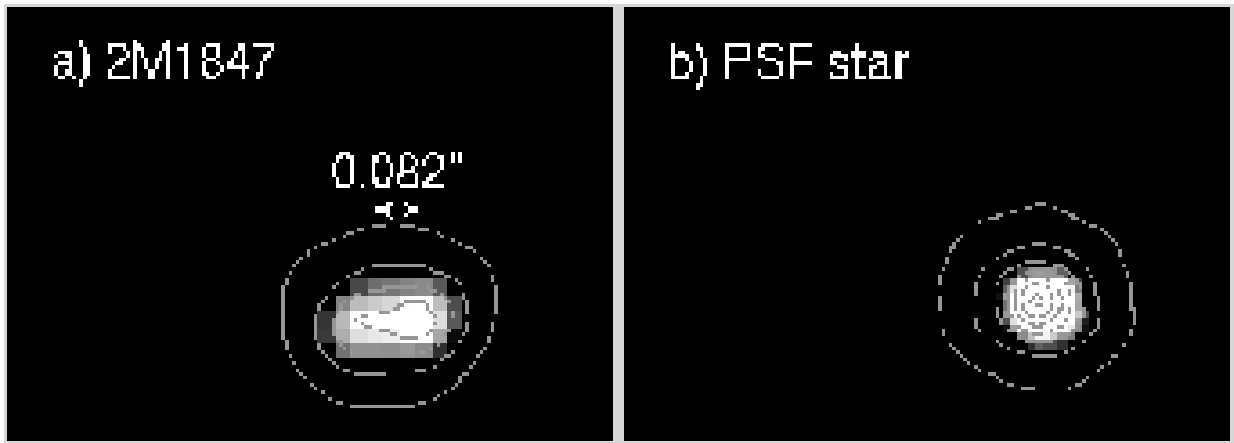}
\caption{(a) A 12$\times$10\,s image of the newly discovered binary
system 2M\,1847 shown in the K$_s$ band; observed on 2003 July 10 (UT)
at Subaru. The platescale is 0$\arcsec$.0217\,pixel$^{-1}$. The
contours are linear at the 80, 60, 40, and 20\% levels. (b) For
comparison, the PSF star 2M\,0253 observed in the same evening and at
the same airmass is
displayed. The contours are linear at the 85, 75, 65, 55, 45, 30, and 15\%
levels. In both images, north is up and east is to the left.
\label{fig8}} 
\end{figure}

The images were reduced using an AO data reduction pipeline written in
the IRAF language as first described in \cite{clo02}. The pipeline
produces final unsaturated exposures in J, H, and K$_s$ with deep
720\,s exposures at H band for each observed binary system. The
dithering of the shorter exposures produces a final
30$\arcsec\times30\arcsec$ image with a high S/N region in a
10$\arcsec\times10\arcsec$ box centered on the binary. In order to
detect close companions within 1$\arcsec$ of the central star we
filter out the low spatial frequency components of the deep images
leaving behind high frequency residuals in the PSF (unsharp
masking). No faint companions, however, were found within the halo of
our central stars using this technique. Both binary systems were
detected from reductions of the shorter exposures. Figures
\ref{fig1} and \ref{fig8} show K$_s$ images of the two new systems.

%\placefigure{fig1} 

Photometry for the more widely separated 2M\,0429 was performed using the DAOPHOT PSF fitting photometry package in IRAF. The PSFs used were unsaturated single stars observed during the same night with similar IR brightness, spectral type, and air mass. The errors in $\Delta$mag, listed in Table \ref{tbl-2}, are the differences in the photometry between 2 similar PSF stars. 

DAOPHOT could not successfully separate the strongly blended
components of 2M\,1847 due to lower Strehl ratios caused by observing
through a 1.4 airmass (the Strehl ratio and hence resolution were
better at airmass of 1 when the binary was initially discovered and
its components more clearly separated, however,
technical difficulties resulted in delayed image acquisition). Consequently,
we remove the low spatial frequencies of the binary revealing their
high-frequency cores. We then perform aperture photometry using IRAF
PHOT. This purely differential technique preserves the relative
magnitude difference between each component while removing sufficient
primary halo flux to reveal the companion centroids in K$_s$. The
technique gave reliable $\Delta$mags and was verified on binary images
with known $\Delta$mags. 

We calculate individual fluxes and their uncertainties from the measured binary flux ratios and the integrated 2MASS apparent magnitudes (2MASS All-Sky Point Source Catalog), along with their respective uncertainties. Table \ref{tbl-3} lists the photometry and derived characteristics of the new binary systems. 

%\placetable{tbl-3}

\subsection{An Example of Sensitivity: The Special Case of 2MASS\,0253} \label{bozomath_e}
One of our targets observed to have no stellar companions deserves special mention. 2M02530084+1652532 (hereafter 2M\,0253) is a newly discovered M6.5 dwarf \citep{tee03} remarkably only 3.6\,pc away. It was discovered from a search of the SkyMorph database of the Near Earth Asteroid Tracking (NEAT) project \citep{pra99} as a high proper motion object (5$\arcsec$yr$^{-1}$). The star's proximity presented a rare observational window for the direct imaging of several Jupiter-mass, extrasolar planets. We were able to both probe semimajor axis separations to within $\sim$3\,AU of the star, comparable to the separations of known extrasolar planets detected through radial velocity studies (a\,$\lesssim6$\,AU; http://exoplanets.org) {\it and} outside the speckle-dominated region on the detector ($\gtrsim1\arcsec$). We were the first to observe this object with high resolution on 2003 July 14 (UT) using the NIRC2 camera and AO \citep{wiz00} on the W. M. Keck II telescope. The 0$\farcs01\,$pixel$^{-1}$ plate scale mode was used on the 1024$\times$1024 pixel array.

2M\,0253 was only observable for approximately 1 hour in the early morning. We achieved sensitivity to companions of H=19.6 at 1$\farcs5$ in 24\,min of total integration time (49$\times$30\,s frames in a 4-dither pattern). We fully saturated the central star so as to allow for the detection of any massive faint Jupiter planets orbiting $\ga1\arcsec\, (>2.6\,$AU) from the central star. No faint companions were detected. 

\begin{figure}
\includegraphics[angle=90,width=\columnwidth]{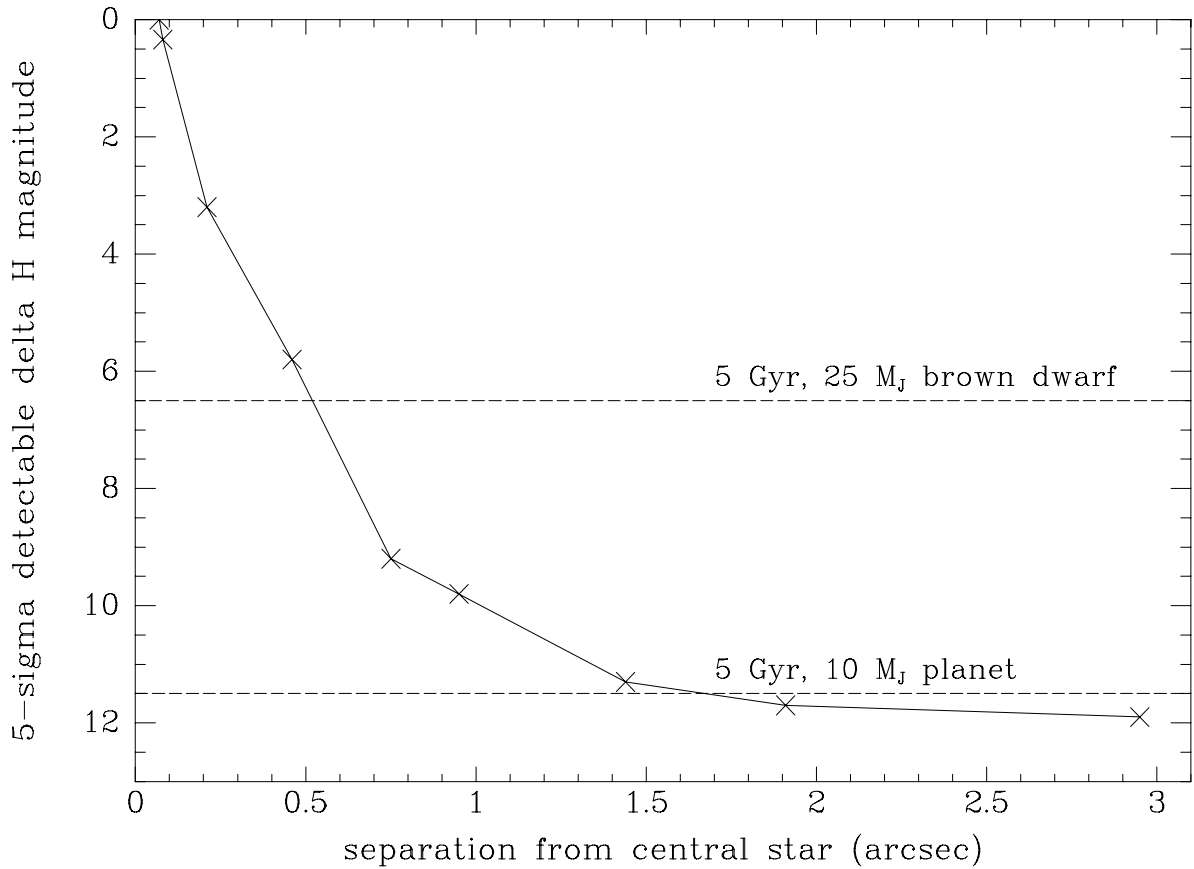}
\caption{Instrument sensitivity curve showing 5$\sigma$ $\Delta$H detection versus distance in arcsec from the very nearby M6.5 star 2M\,0253 . Total integration time is 24\,min using Keck II AO. The ``crosses'' indicate the 5$\sigma$ sensitivity limits of our data to simulated faint companions. The upper horizontal dashed line corresponds to a 5\,Gyr old, 25 M$_{J}$ brown dwarf using the models of \cite{bur03} while the lower horizontal dashed line corresponds to a 5\,Gyr old, 10 M$_{J}$ planet. We were sensitive to the detection of a 10 M$_{J}$ planet at 4\,AU and a 25 M$_{J}$ brown dwarf at 1.5\,AU (assuming 2M\,0253 is only 3.6\,pc away and 5\,Gyr old).
\label{fig2}} 
%sensitivity.eps=f2.eps
\end{figure}

\begin{figure}
\includegraphics[angle=0,width=\columnwidth]{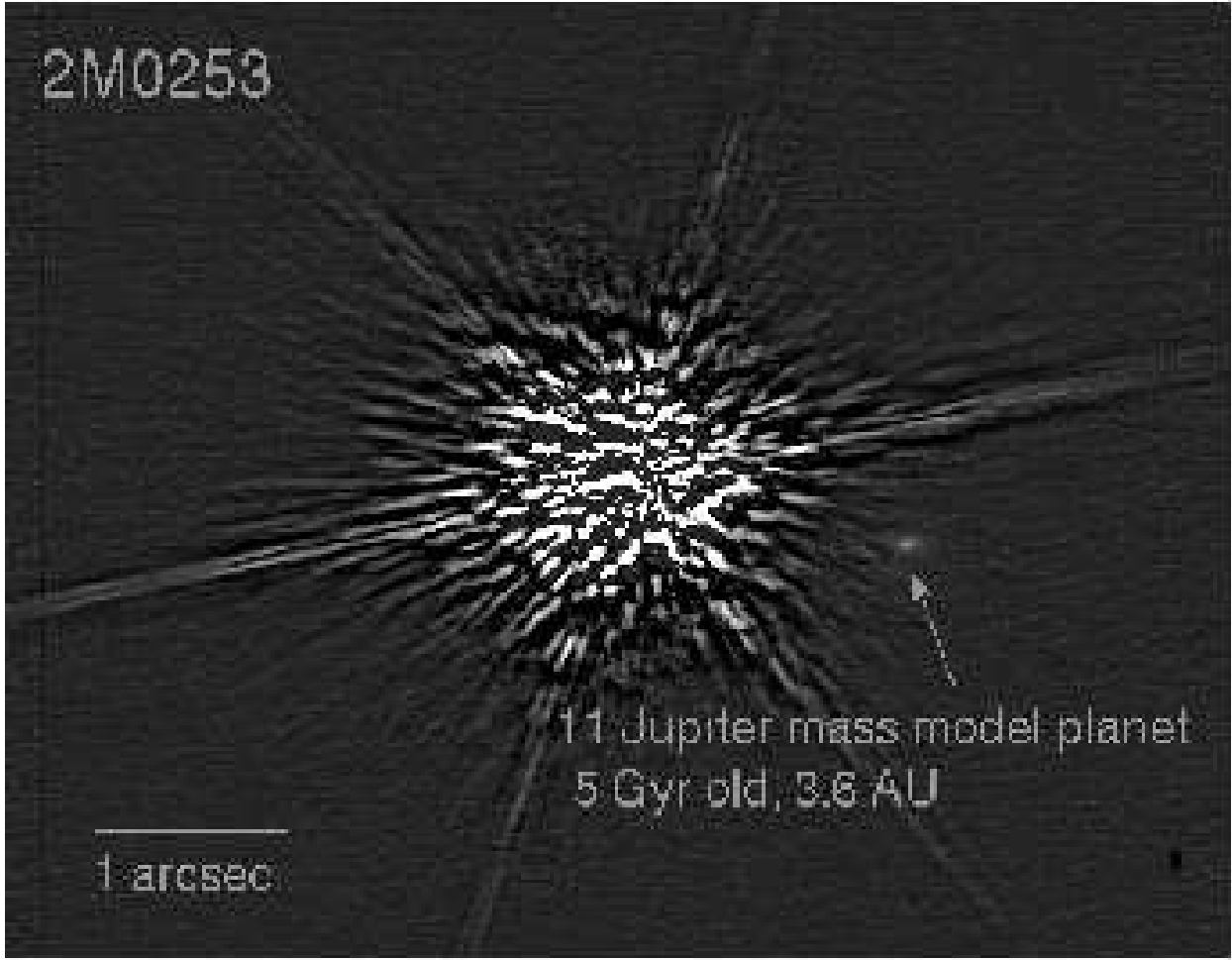}
\caption{Reduced 24\,min image of 2M\,0253 observed at Keck II on 2003 July 14 (UT) with its low spatial frequencies removed (unsharp masked). The simulated 5$\sigma$ companion is 33,000 times fainter than the central star ($\Delta$H=11.3) at only 3.6\, AU (1.5$\arcsec$). \cite{tee03} report H=7.9 for the central star which allows detection to a 5\,Gyr old, 11\,M$_{J}$ planet according to the models of \cite{bur03}. Also visible are the residuals from the 6 spider arms and super speckles. To within our sensitivity limits we find no companions to  2M\,0253.
\label{fig3}} 
%2M0253_planet.eps=f3.eps
\end{figure}

The 24\,min of total integration time enables us to establish upper limits on planetary masses orbiting this star. We construct an unsaturated PSF of the star by replacing the saturated core with scaled unsaturated pixels from a short exposure. We determine maximum H band $\Delta$\,mag contrasts by combining scaled models of faint companions (with appropriate PSFs) at various radial distances until a 5$\sigma$ detection is obtained. Figure \ref{fig2} shows the resulting 5$\sigma$ limiting magnitudes at several radial distances from the central star. The horizontal dashed lines indicate the H band $\Delta$\,mag required for the detection of 5\,Gyr, 10\,M$_J$ and 25\,M$_J$ objects using the models of \cite{bur03}. We use the peak of their H band spectra to estimate the flux emission in this exercise. The star's age is not known but based on its high tangential velocity we can assume it is an older object \citep{wie74}. The figure demonstrates sensitivity to an 11 Jupiter-mass extrasolar planet at only $\sim\,$4\,AU (1.5$\arcsec$) away. In Figure \ref{fig3} we show the fully reduced Keck image of 2M\,0253 spatially filtered of its low frequency halo, leaving behind high frequency residuals in the core (superspeckles). This image also illustrates that with conventional AO, speckle noise limits the detection of faint companions within the inner $\sim1\arcsec$ of the halo. 

\section{Analysis}
\subsection{Are the Companions Physically Related to the Primaries?} \label{bozomath_f}
From the total of 69 objects already observed in both this survey and
a companion survey of M8.0-L0.5 stars by the authors \citep{clo04}, we
did not detect any additional unknown red (J\,-\,K$_s>0.8$\,mag)
background objects in $6.2\times10^4$ square arcsec. Therefore, we
estimate the probability of a chance projection of a comparably red
object within 0.5$\arcsec$ of the primary to be $<1.3\times
10^{-5}$. As we argued in Paper I, with an M6-M8 dwarf density of
0.007\,pc$^{-3}$ in the local solar neighborhood \citep{rei97}, the
probability of an apparent companion being just a background star at,
for example, twice the distance of the target star (hence fainter by a
$\Delta$ magnitude of 1.5 mag) {\it and} appearing within $0.5\arcsec$
of any of our targets is estimated to only be
$\sim3\times10^{-7}$. Additionally, none of the companion images
appear spatially extended as might be expected of background
galaxies. Therefore, we conclude that both of the very red companions
are physically associated with their primaries and hereafter we will refer to them as 2M\,0429B and 2M\,1847B.

\subsection{Distances} \label{bozomath_g}
Neither of the 2 binary systems have published trigonometric
parallaxes. We estimate distances to both primaries from a
color-magnitude diagram developed in Paper I based on trigonometric
parallaxes of other well-studied, late-M, field dwarfs from \cite{dah02}. Using corresponding 2MASS photometry for each star with a trigonometric parallax, we estimated a linear least-squares fit of M$_{K_s} = 7.65 + 2.13\,$(J-K$_s)$ for the spectral range M6.5-L1. This relationship has a 1$\sigma$ error of 0.33 mag, which has been added in quadrature to the J and K$_s$ photometric errors to yield the primarys' M$_{K_s}$ values reported in Table \ref{tbl-3}. We then use the distance modulus of the primary to estimate the distances to the binaries. The calculated distances are listed in Table \ref{tbl-3}. 

\subsection{Spectral Types and Temperatures} \label{bozomath_h}
We do not have spatially resolved spectra of the individual components in either of the 2 new systems. We estimate the spectral types of each of the binary components by using the relation SpT = 3.54M$_{K_s}$ - 27.20 derived in Paper I from the data set of \cite{dah02} (eg. SpT\,=\,8 is an M8, SpT\,=\,10 is an L0, etc). This relationship has a 1\,$\sigma$ error of 0.85 spectral types which when taken in quadrature with the uncertainty in M$_{K_s}$ gives an overall uncertainty of about 1.5 spectral types. Fortunately, none of analysis is dependent on these spectral type estimates. The results are listed in Table \ref{tbl-3}. 

\begin{figure}
\includegraphics[angle=0,width=\columnwidth]{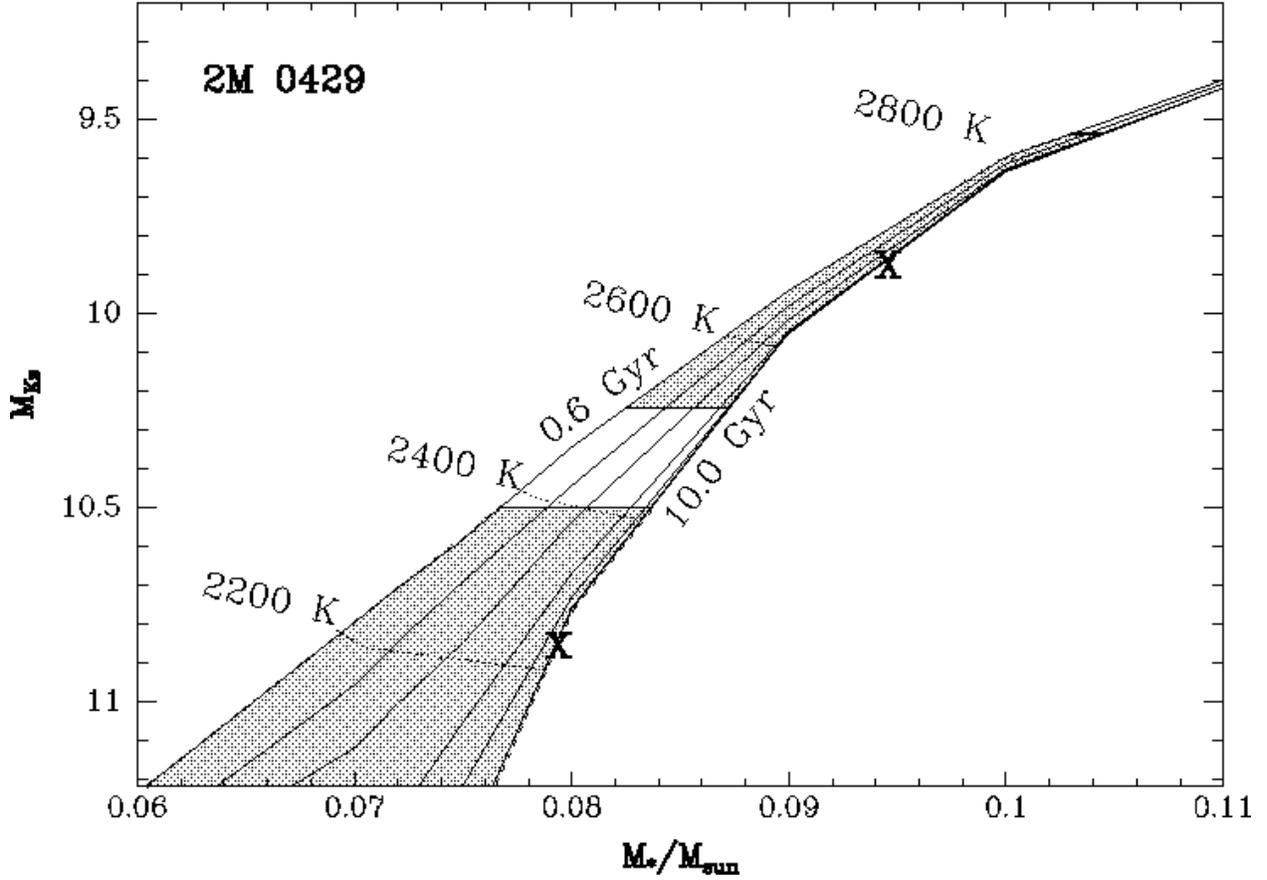}
\caption{\cite{cha00} DUSTY stellar and substellar evolutionary tracks custom integrated over the K$_s$ bandpass ([{\it m}/H]\,=\,0). The best-guess values of the individual binary components of 2M\,0429 are indicated by the bold ``crosses'' with the primary at the top right and the companion lower and to the left. The shaded polygons enclose each components' region of uncertainty. The components' derived M$_{K_s}$ is listed in Table \ref{tbl-3}. With no knowledge of the binary's age, we conservatively assign a mean age of 5\,Gyr and uncertainties spanning the range of ages in the solar neighborhood \citep[0.6\,-\,7.5 Gyr;][]{cal99}. The model suggests a primary mass of 0.094$_{-0.011}^{+0.010}$\,M$_{\sun}$ and a temperature of 2690$_{-170}^{+160}$\,K. For the companion, the model predicts a mass of 0.079$_{-0.018}^{+0.005}$\,M$_{\sun}$ and a temperature of 2240$_{-260}^{+190}$\,K. The isochrones plotted are (left to right) 0.6, 0.65, 0.7, 0.85, 1.2, 1.7, 3.0, 5.0, 7.5, and 10.0 Gyr (the oldest 4 isochrones are indistinguishable at the given scaling). 
\label{fig4}} 
%2M0429smaller.eps=f4.eps
\end{figure}

\begin{figure}
\includegraphics[angle=0,width=\columnwidth]{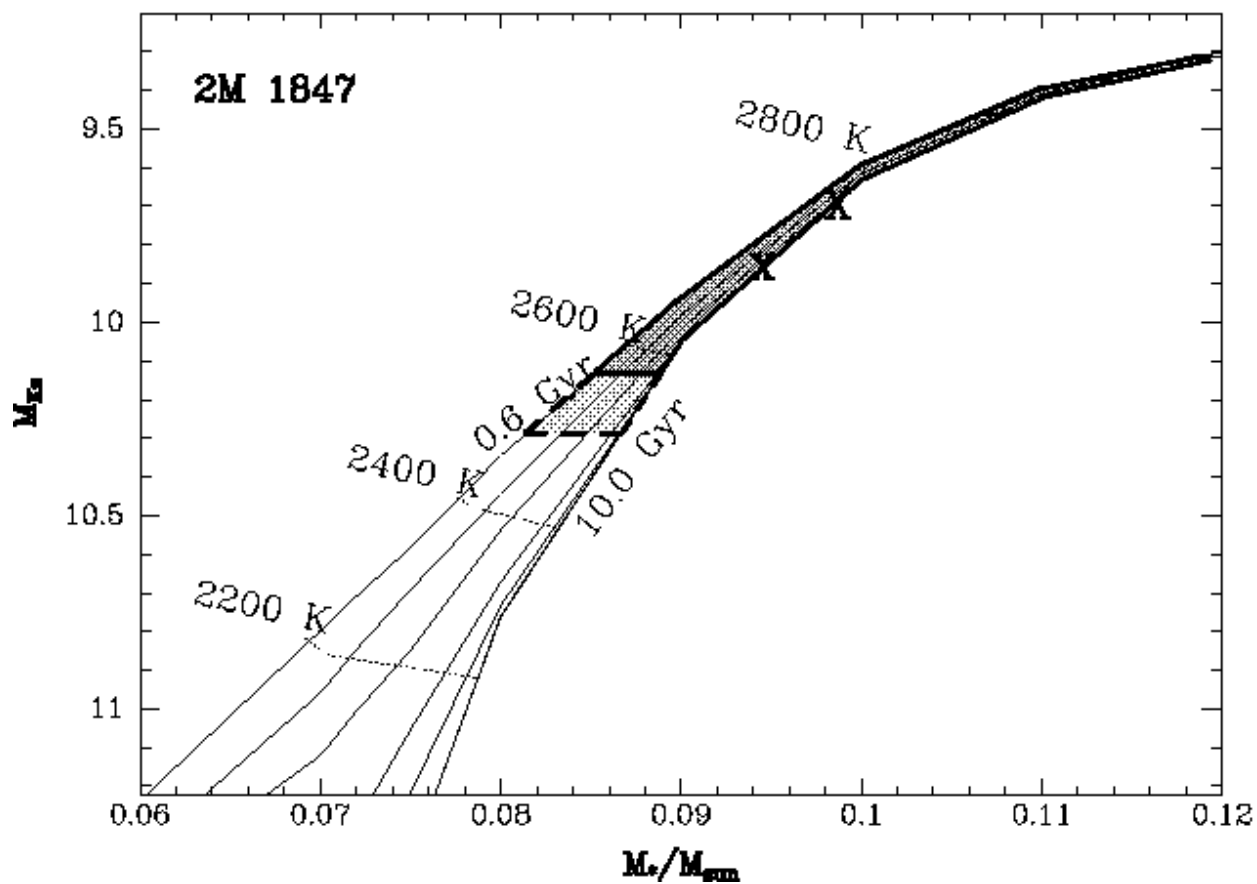}
\caption{As in Figure \ref{fig4}, but for 2M\,1847. In this case the shaded regions of uncertainty of the 2 components overlap. Hence, we outline the primary star's in solid and the companion in dotted lines (where the two regions overlap solid takes preference). The model suggests a primary mass of 0.098$_{-0.012}^{+0.022}$\,M$_{\sun}$ and a temperature of 2760$_{-260}^{+280}$\,K. For the secondary the model suggests a mass of 0.094$_{-0.013}^{+0.014}$\,M$_{\sun}$ and temperature of 2690$_{-210}^{+220}$\,K.
\label{fig5}} 
%2M1847_final.eps=f5.eps
\end{figure}

Effective temperatures of the binary components are estimated from the DUSTY evolutionary tracks \citep{cha00} using calculated M$_{K_s}$ values and estimated ages (see Figures \ref{fig4} and \ref{fig5}). We estimate 2M\,0429A and 2M\,0429B to have effective temperatures of 2690$_{-170}^{+160}$\,K and 2240$_{-260}^{+190}$\,K, respectively; 2M\,1847A and 2M\,1847B are estimated at 2760$_{-260}^{+280}$\,K and 2690$_{-210}^{+220}$\,K, respectively. These estimated temperatures are in very good agreement with those predicted in \cite{dah02} for the given spectral types.

\subsection{Ages and Masses} \label{bozomath_i}
Estimating the age of late-type field dwarfs without Li measurements or established cluster membership is difficult. Consequently, we conservatively assume a mean age of $\sim5$ Gyr for our objects with uncertainty spanning the range of common ages in the solar neighborhood \citep[0.6\,-\,7.5 Gyr;][]{cal99}.

To estimate masses of these objects we rely on luminosity-mass-age
models for VLM stars and brown dwarfs. We utilize the DUSTY models to
provide theoretical estimates for both stellar and substellar masses
as a function of both absolute K$_s$ magnitude and age. The tracks are
calibrated for the K$_s$ bandpass (I. Baraffe, private communication)
and we extrapolate the isochrones from 0.10 to 0.11\,M$_\sun$ so as to
enclose the upper mass limits of our central stars. The companion's
absolute magnitude is simply determined by adding the measured
$\Delta$K$_s$ to its primary star's M$_{K_s}$. The crosses in Figures
\ref{fig4} and \ref{fig5} indicate the best estimates of where the
binary components lie on the 5\,Gyr tracks and their uncertainties are
represented by the shaded regions. 2M\,0429A's region of uncertainty
is displayed in the upper right while its companion is displayed in
the lower left. Because the 2M\,1847 binary system is of near equal
magnitudes, their regions of uncertainty largely overlap. In this case
the slightly more massive primary's region of uncertainty is indicated
in bold outline and the portion of the companion's not overlapping is
dashed. The maximum mass is related to the minimum M$_{K_s}$ at the
oldest possible age; the minimum mass is related to the maximum
M$_{K_s}$ at the youngest possible age. Table \ref{tbl-3} lists the
estimated masses for both binary systems. Both systems' primary masses
are consistent with M7-type dwarfs and their secondaries are most
likely stellar, however, 2M\,0429B's uncertainties extend well into
the substellar region according to the model. The uncertainty in the masses, as well as in the effective temperatures, is largely driven by the uncertainty in our determination of M$_{K_s}$ ($\sigma$\,=\,0.33 mag) as obtained from the [M$_{K_s}$,\,J-K${_s}$] color magnitude diagram linear fit from Paper I. Future observations of trigonometric parallaxes would significantly reduce the uncertainty in M$_{K_s}$ and hence the masses and temperatures.

\section{Discussion}

\subsection{The Binary Frequency of M6.0-M7.5 Stars}\label{bozomath_j}
We update the binary fraction statistics of M6.0-M7.5 stars combining the latest results presented here (2 binaries resolved out of 6) with those from Paper I (3 binaries out of 30). This implies an observed, uncorrected binary fraction of 14$_{-4}^{+8}\%$ using a Poisson distribution for the uncertainty \citep{ber03}. However, this sample was originally drawn from a magnitude-limited sample and hence the observed binary fraction is biased due to the leakage of equal magnitude binaries into our sample from further distances (Malmquist bias). We therefore need to correct our result due to this bias as well as consider sample incompleteness due to undetected very tight lower mass companions.

To compensate for the fainter single stars not included in our
flux-limited sample we first adjust for a larger observed volume due
to the discovered binaries by a volume correction factor. This factor
is simply the ratio of the spherical volume containing 95\% of our
detected binaries and the spherical volume containing 95\% of our
target objects. This results in a volume correction factor of
(30\,pc/24\,pc)$^3$\,=\,2 and a Malmquist-corrected binary fraction of $5/(36\times2)$ or 7$_{-2}^{+4}\%$.

\begin{figure}
\includegraphics[angle=90,width=\columnwidth]{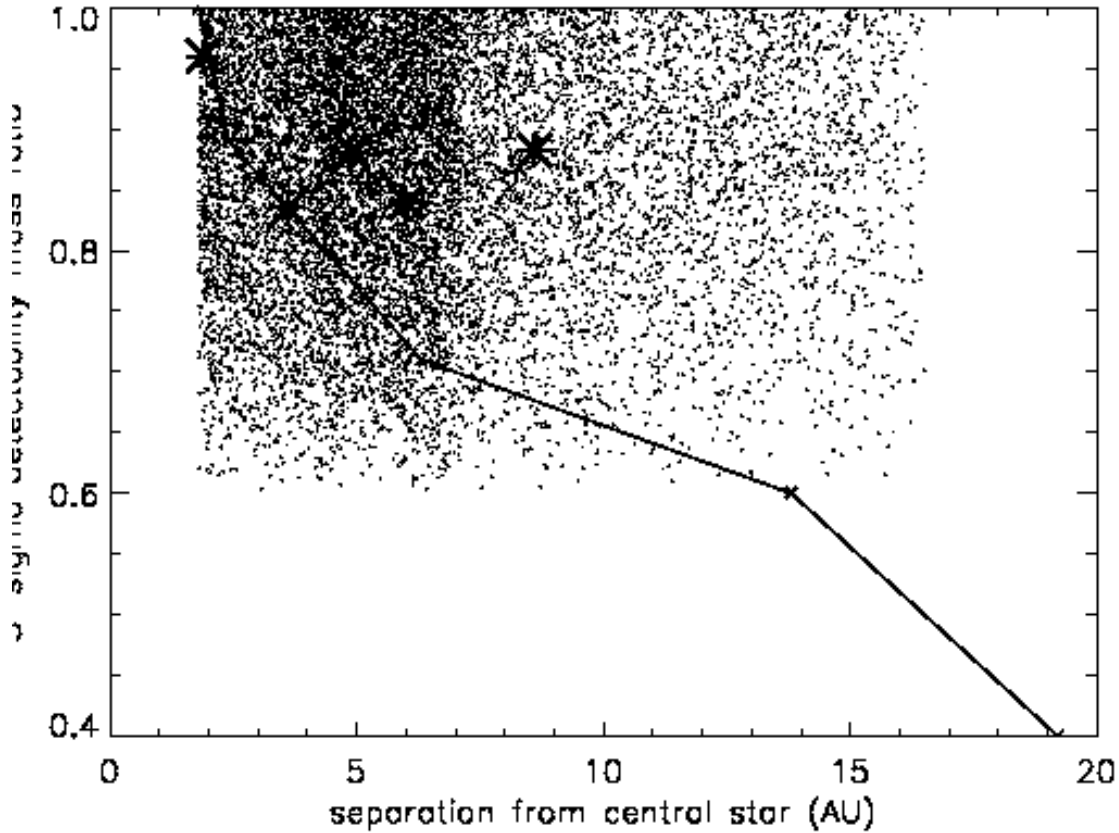}
\caption{The results from a Monte Carlo simulation generating 11,670
companions distributed according to a bivariate distribution (mass
ratio q and separation a) is plotted over the instrumentation
sensitivity curve (connected lines). We assume the two distributions
are independent. We assume a power law declining from unity to 0.6 for
the mass ratio distribution \citep{clo04} and the profile from Figure
\ref{fig7} for the separation distribution for a\,$\,>3$\,AU. The
instrumentation sensitivity curve is based on modeling of a
$\sim5\,$Gyr M6.5 dwarf placed at 30\,pc, typical of the distances of
our discovered binaries. The DUSTY models \citep{cha00} are used to
convert $\Delta$H magnitudes to mass ratios. The 5 discovered binary
systems are indicated by large asterisks. 21$\%$ of the synthetic
companions fall below the instrumentation sensitivity curve but above
the instrument sensitivity mass ratio cutoff of q=0.6. This results in
a sensitivity correction of 1.3 binaries.   
\label{fig6}} 
%nickplot.eps=f6.eps
\end{figure}

The possibility that there were faint companions, both stellar and
non-stellar, not detected due to instrument insensitivity is a real
one. The curve in Figure \ref{fig6} shows the instrumental sensitivity
of our sample in the speckle noise limited region ($<1\arcsec$, 30\,AU
for a star assumed 30\,pc away) as a function of mass ratio and
projected separation in AU. It is based on modelling of a 5\,Gyr
(typical of the ages expected for field stars \cite{giz00}), M6.5
dwarf placed at 30\,pc (typical of the distances of our discovered
binaries) observed at the 8-m Gemini North telescope. We use the DUSTY
models to convert $\Delta$H magnitudes to mass ratios. The reason we convert to mass ratios is
because it allows us to use the observed mass ratio distribution for
VLM binaries \citep{clo04} to predict the number of missed companions
with q\,$\ge0.6$. The 5 large asterisks in Figure \ref{fig6} represent the 5 discovered binary systems from this survey. Interestingly, they are all found in the upper left corner of the sensitivity curve. The fact that some are so near the curve strongly infers that binaries just below the sensitivity curve were most likely missed. 

\begin{figure}
\includegraphics[angle=90,width=\columnwidth]{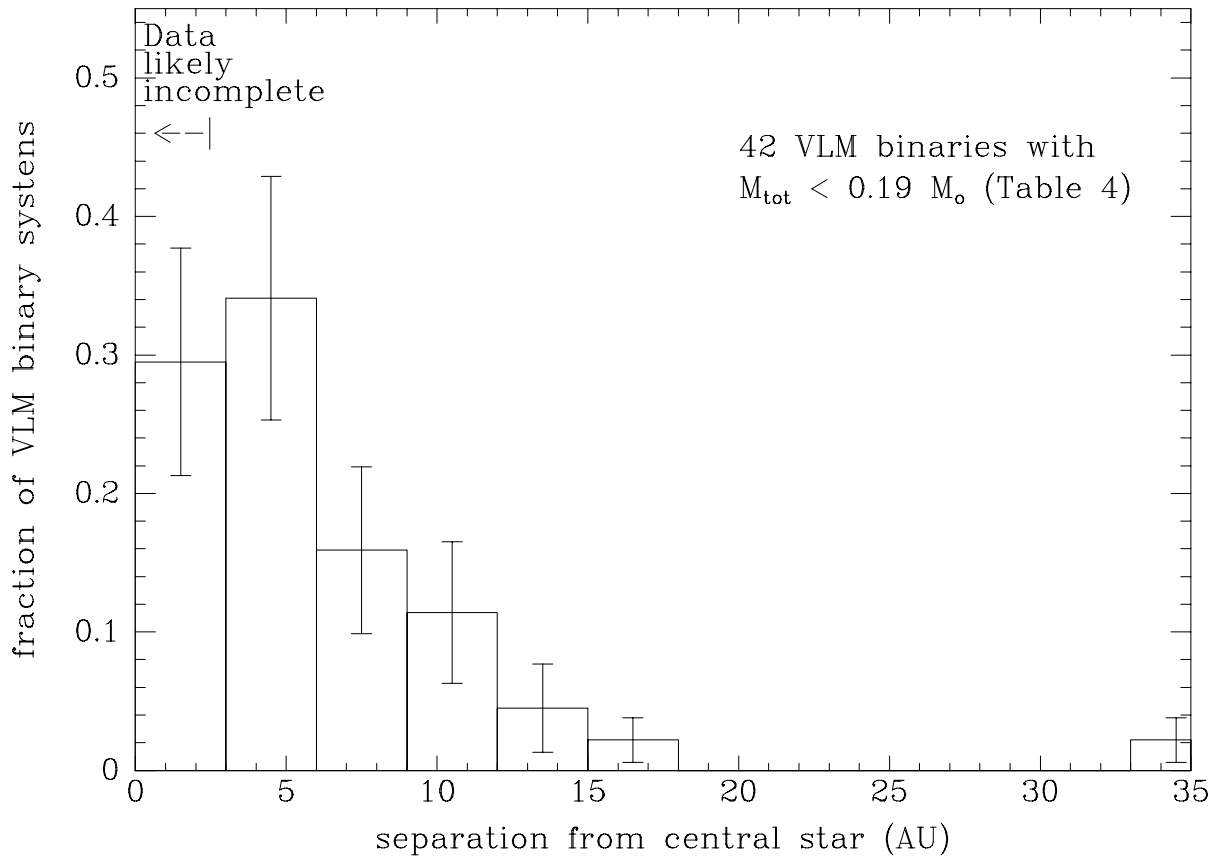}
\caption{A histogram of 42 VLM (M$_{tot}\,<\,0.19\,$M$_\sun$,
component spectral types $\geq$\,M6.0) binaries from Table 4 are
plotted (for reasons of clarity, we leave out the 2 widest systems). The distribution is incomplete less than $\sim3\,$AU; Poisson error bars are plotted. The declining profile is a real feature of the distribution as is the paucity of wide binaries greater than 15\,AU. These features along with a tighter peak distribution ($\sim3-4\,$AU) are significantly different from those of more massive primary stars \citep{fis92, duq91}.
\label{fig7}} 
%sep_dist.eps=f7.eps
\end{figure}

To apply an instrument-sensitivity correction we need to estimate how
many binaries went undetected in our sample. We generate a Monte Carlo
simulation of 11670 synthetic companions with the binary properties of
VLM systems. For our model we use the mass ratio and separation
distributions for VLM binaries and assume that the distributions are
independent. For the mass ratio distribution we assume a power law
decline from unity to 0.6 from \cite{clo04}. We create the separation
distribution profile by plotting the 42 most tightly
separated and resolved VLM (M$_{tot}\,<\,0.19\,$M$_\sun$) binaries
currently known (see Figure \ref{fig7}; Table 4). Originally presented
in \cite{clo04}, we update the list of all known VLM binaries and
present it in Table 4. The definition of VLM binary having a total
mass of $<$0.19\,M$_\sun$ is selected to ensure that the binary
components are of spectral type M6.0 or later and hence differentiated
from more massive systems. The peak of this distribution, $\sim5$\,AU,
is much tighter than the $\sim30$\,AU distribution peak of slightly
more massive M0-M4 dwarfs \citep{fis92} and solar-mass stars
\citep{duq91}. The separation distribution is bound by the smallest
separation the instruments were able to obtain in H band
($\sim0\farcs08\times30$\,pc) on the near side and the empirically sampled
wider separation encompassing 95\% of known VLM binaries (Table 4) on the far side. 

From this sample of nearly 12,000 simulated companions, 21\% were below
the instrument sensitivity curve (but above the instrument sensitivity
mass ratio cutoff of q\,=\,0.6) as shown in Figure \ref{fig6}. With 5
detected binaries, this predicts 1.3 companions were missed in our
survey. Hence we conclude that the binary fraction for M6.0-M7.5 stars
is (5+1.3)/36/2 or 9$_{-3}^{+4}\%$. It should
be pointed out, however, that the true fraction is most certainly
larger than this figure since we cannot rule out the possibility of
the existence of low q binaries due to the sensitivity of this
survey. Therefore, our reported binary fraction, accurate within the
uncertainties for M6.0-M7.5 dwarfs for separations $\ga$ 3\,AU, represents
a low-end estimate to the intrinsic binary fraction.

With slightly improved statistics, this latest result for the binary fraction of M6.0-M7.5 stars is now more comparable with those of later spectral types: 15$\pm7\%$ for late M/early L dwarfs \citep{clo04}, 10\,-\,15$\%$ and 15$\pm5\%$ for L dwarfs \cite[respectively]{bou03,giz03} and 9$_{-4}^{+15}\%$ for T dwarfs \citep{ber03}. These cooler dwarfs including the ones presented here all have binary fractions significantly lower than the $\sim32\%$ observed for earlier M0-M4 dwarfs \citep{fis92} and the $\sim50\%$ for solar-mass stars \citep{duq91} over the same a\,$>$\,3\,AU separation range. Our conclusion from Paper I is strengthened: {\it for spectral type M6.0-M7.5 binary systems with separations 3\,AU\,$<$\,a\,$<\,$300\,AU the binary fraction from our survey is 9$_{-3}^{+4}\%$, statistically consistent with cooler M, L, and T stars and significantly less common than that of G and early M stars.} 

\subsection{The Separation Distribution Function}\label{bozomath_k}
The 2M\,0429 and 2M\,1847 binary systems have best-estimate projected
separations of 6\,AU and 2\,AU, respectively. In our total M6.0-M7.5
sample, we detect no binary separations wider than 10\,AU (Table
3). When we analyze the semimajor axis separations of these binaries
we observe that there are no projected VLM binary separations
$>15$\,AU. With our survey sensitive out to $\sim15\arcsec$ from the
central star, this result appears to be real and
not a result of a sensitivity selection effect. When examining the
entire VLM sample of known binary systems from the literature (44,
Table 4), only three objects are currently known to have a projected
separation $>15$\,AU. This indicates that while wide VLM binaries of
q\,$>$\,0.6 can exist, they are rare.

The median projected separation of our entire binary sample is $\sim5$\,AU, consistent with the $\sim$\,4 AU peak distribution of late M/early L binaries \citep{clo04}, L dwarfs \citep{giz03, bou03}, and T dwarfs \citep{ber03}. This contrasts significantly with the $\sim30$\ AU broad separation peak of early M and G dwarfs \citep{fis92,duq91}. {\it We conclude that the projected semimajor axes of M6.0-M7.5 binaries appear consistent with those of late M, L, and T dwarf systems but are significantly smaller on average than early M and G stars.}

\section{Summary}\label{bozomath_l}

We have conducted the largest flux limited (K$_s<12$\,mag) survey of nearby M6.0-M7.5 dwarfs using the Keck II, Gemini North, Subaru, and VLT AO systems. In this paper we present our 2 latest binary discoveries, 2M\,0479 and 2M\,1847, observed at the VLT and Subaru facilities, respectively. When added to our initial results from Paper I, the overall survey consists of 36 stars with 5 discovered binary systems. The 2 new components are of relatively equal mass (q$\,>$\,0.8) with average projected separations of 2 and 6\,AU. While none of the binaries have yet been confirmed by common proper motions, they are almost certainly bound based on space density arguments of very red companions. We have used observational and statistical arguments to characterize the VLM binary fraction and separations that contribute additional empirical constraints to binary formation mechanisms:
\begin{itemize}
\item We estimate the binary frequency of spectral type M6.0-M7.5 main sequence stars for separations a\,$>$3\,AU from this survey to be 9$_{-3}^{+4}\%$. The figure is statistically consistent with later type ultracool M, L, and T dwarfs. The frequency is significantly less than that measured in studies of earlier M and G dwarfs, inferring that the binary fraction of stars is a function of the spectral type of the central star.  

\item The separations of the 5 binary systems from our sample are all $<10$ AU. Projected separations of known VLM binaries $>15$\,AU are rare. This survey's median separation of 5\,AU is consistent with the separations of later type M, L, and T dwarfs (separation peak $\sim4$\,AU). This is in stark contrast with the broad peak separations of $\sim30$ AU for the more massive M and G binaries.

\end{itemize}

\acknowledgements
We thank Adam Burgasser for elucidating discussion regarding binary fraction statistics, Chien Peng for suggestions regarding Monte Carlo simulations, and Dan Potter for IDL assistance. K.L.C. acknowledges support from a NSF Graduate Research Fellowship. This publication makes use of data products from the Two-Micron All-Sky Survey, which is a joint project of the University of Massachusetts and the Infrared Processing and Analysis Center/California Institute of Technology, funded by NASA and the NSF.

\clearpage
\begin{deluxetable}{lllcclllll}
\tabletypesize{\scriptsize}
\tablecaption{M6.0-M7.5 Stars Observed with No Likely Physical Companion Detections Between 0.1$\arcsec$-15$\arcsec$\tablenotemark{a}\label{tbl-1}}
\tablewidth{0pt}
\tablehead{
\colhead{2MASS Name} &
\colhead{Other Name} &
\colhead{K$_s$} &
\colhead{Spectral Type} &
%\colhead{Telescope}
\colhead{Reference} &
}
\startdata
2MASS\,\,\,\,\,\,\, J0253008+165253\tablenotemark{b}&SO 025300.5+165258 & \,\,\,7.59 & M6.5 & 5\\
2MASSI\,\,\,\, J0330050+240528 &LP 356-770& 11.36 & M7.0 & 1\\
2MASSI\,\,\,\, J0435161-160657\tablenotemark{b}&LP775-31&\,\,\,9.34&M7.0& 2\\
2MASSI\,\,\,\, J0752239+161215 && \,\,\,9.82 & M7.0 & 2\\
2MASSI\,\,\,\, J0818580+233352 && 11.13 & M7.0 &  1\\
2MASSW J0952219-192431 && 10.85 & M7.0 & 1\\
2MASSW J1016347+275150 & LHS 2243 & 10.95 & M7.5 & 1\\
2MASSI\,\,\,\, J1024099+181553 && 11.21 & M7.0 & 1\\
2MASSW J1049414+253852 && 11.39 & M6.0 & 1\\
2MASSI\,\,\,\, J1124532+132253 && 10.03 & M6.5 & 2\\
2MASSW J1200329+204851 && 11.82 & M7.0 & 1\\
2MASSW J1237270-211748 && 11.64 & M6.0 & 1\\
2MASSW J1246517+314811 & LHS 2632 & 11.23 & M6.5 & 1\\
2MASSI\,\,\,\, J1253124+403404 && 11.20 & M7.5 & 4\\
2MASSW J1336504+475131 && 11.63 & M7.0 & 1\\
2MASSW J1344582+771551 && 11.83 & M7.0 & 1\\
2MASSI\,\,\,\, J1356414+434258 && 10.63 & M7.5 & 2\\
2MASSI\,\,\,\, J1431304+171758\tablenotemark{b} && 11.16 & M6.5 & 2\\
2MASSI\,\,\,\, J1521010+505323\tablenotemark{b} && 10.92 & M7.5 & 2\\
2MASSP\,\, J1524248+292535 && 10.15 & M7.5 & 3\\
2MASSW J1527194+413047 && 11.47 & M7.5 & 3\\
2MASSW J1543581+320642 &LP 328-36& 11.73 & M7.5 & 1\\
2MASSW J1546054+374946 & &11.42 & M7.5 & 1\\
2MASSW J1550381+304103 && 11.92 & M7.5 & 1\\
2MASSW J1757154+704201 &LP 44-162& 10.37 & M7.5 & 1\\
2MASSW J2052086-231809 &LP 872-22& 11.26 & M6.5 & 1\\
2MASSW J2221544+272907 && 11.52 & M6.0 & 1\\
2MASSW J2233478+354747 &LP 288-31& 11.88 & M6.0 & 1\\
2MASSI\,\,\,\, J2235490+184029 &LP 460-44& 11.33 & M7.0 & 1\\
2MASSW J2306292-050227 && 10.29 & M7.5 & 1\\
2MASSW J2313472+211729 &LP 461-11& 10.42 & M6.0 & 1\\
\enddata
\tablenotetext{a}{For near-equal mass companions. For smaller companion masses with q\,$<$0.8, sensitivity is a function of distance. See Figures \ref{fig2} and \ref{fig6}.}
\tablenotetext{b}{Results from this paper; otherwise, Paper I.}
\tablerefs{
(1) \cite{giz00} (2) \cite{cru03} (3) \cite{rei02} (4) \cite{kir91} (5) \cite{tee03}.
}
\end{deluxetable}

\clearpage
\begin{deluxetable}{llllllllll}
\tabletypesize{\tiny}
\tablecaption{The New Binary Systems \label{tbl-2}}
\tablewidth{0pt}
\tablehead{
\colhead{System} &
\colhead{$\Delta J$} &
\colhead{$\Delta H$} &
\colhead{$\Delta K_s$} &
\colhead{Sep. (mas)} &
\colhead{P.A. (deg)} &
\colhead{Date Observed (UT) }&
\colhead{Telescope} &
}
\startdata
LP\,415-20\tablenotemark{a} & $0.84\pm0.15$ & $0.77\pm0.10$ & $0.66\pm0.06$ & $119\pm8$ &\, $91.2\pm0.7$ & 2002 Feb. 07 & Gemini North \\
LP\,475-855\tablenotemark{b} & $0.48\pm0.05$ & $0.43\pm0.04$ & $0.48\pm0.03$ & $294\pm5$ & $131.6\pm0.5$ & 2001 Sep. 22 & Gemini North \\
2MASSI \,\,J0429184-312356\tablenotemark{c} & $1.20\pm0.12$ & $1.10\pm0.08$& $0.98\pm0.08$&$531\pm2$&$298.9\pm0.2$&2003 Feb. 13 & VLT\\
2MASSW J1750129+442404 & $0.74\pm0.15$ & $0.73\pm0.15$ & $0.64\pm0.10$ & $158\pm5$ & $339.6\pm0.7$ & 2002 Apr. 25 & Gemini North\\
2MASSI\,\,J1847034+552243\tablenotemark{c} & $0.26\pm0.18$ & $0.34\pm0.15$ & $0.16\pm0.10$ &\, $82\pm5$ & \,\,\,\,\,91.1$\pm1.4$ & 2003 July \,10 & Subaru\\
\enddata
\tablenotetext{a}{Also known as Bryja\,262.} 
\tablenotetext{b}{Also known as [LHD94]\,042614.2+13312 and 2MASSW\,J0429028+133759.}
\tablenotetext{c}{Results from this paper; otherwise, Paper I.}
\end{deluxetable}

\clearpage
\begin{deluxetable}{llllllllll}
\tabletypesize{\tiny}
\tablecaption{Summary of the Binaries' A and B Components\label{tbl-3}}
\tablewidth{0pt}
\tablehead{
\colhead{Name} &
\colhead{$J$} &
\colhead{$H$} &
\colhead{$K_s$} &
\colhead{$M_{K_s}$\tablenotemark{a}} &
\colhead{SpT\tablenotemark{b}}&
\colhead{d$_{phot}$ (pc)\tablenotemark{c}} &
\colhead{Mass (M$_{\sun}$)\tablenotemark{d}} &
\colhead{Sep. (AU)\tablenotemark{e}} &
\colhead{P (yr)\tablenotemark{f}} 
}
\startdata
LP\,415-20A & $13.09\pm 0.06$ & $12.47\pm 0.05$ & $12.12\pm 0.04$ &\phn $9.72\pm 0.38$ & M7.0 & $30\pm5$ & $0.097_{-0.012}^{+0.011}$ & $3.6\pm 0.7$ & $23_{-6}^{+7}$ \\
LP\,415-20B &$13.93\pm 0.16$ &$13.24\pm 0.11$ &$12.78\pm 0.08$ & $10.37\pm0.39$&M9.5 & &$0.081_{-0.010}^{+0.009}$ \\

LP\,475-855A & $13.21\pm 0.04$ & $12.54\pm 0.04$ & $12.18\pm 0.04$ &\phn $9.84\pm0.36$ & M7.5 & $29\pm5$ & $0.093_{-0.009}^{+0.012}$ &  $8.6\pm 1.5$ & $86_{-19}^{+20}$ \\
LP\,475-855B&$13.69\pm 0.07$ & $12.97\pm 0.06$ &$12.66\pm 0.05$ & $10.32 \pm 0.36$&M9.5&& $0.082_{-0.009}^{+0.009}$  \\

2M\,0429A\tablenotemark{g} & $11.18\pm 0.04$ & $10.55 \pm 0.03$ & $10.14 \pm 0.03$ &\phn $9.88\pm 0.35$ & M7.5 & $11\pm2$ & $0.094_{-0.011}^{+0.010}$ & $6.0\pm 1.0$ & $50_{-11}^{+12}$ \\
2M\,0429B\tablenotemark{g} & $12.38\pm 0.13$ & $11.65 \pm 0.09$ & $11.12 \pm 0.07$ &\phn $10.86\pm 0.36$ & L1.0 & & $0.079_{-0.018}^{+0.005}$      \\

2M\,1750A & $13.23\pm 0.06$ & $12.62\pm 0.06$ & $12.24\pm 0.05$ &\phn $9.77\pm0.39$ & M7.5 & $31\pm6$ & $0.097_{-0.012}^{+0.012}$ &  $4.9\pm 0.9$ & $36_{-9}^{+10}$ \\
2M1750B & $13.97\pm 0.16$ & $13.35\pm 0.16$ &$12.88\pm 0.11$ & $10.41 \pm 0.41$ & M9.5 && $0.085_{-0.016}^{+0.006}$ \\

2M\,1847A\tablenotemark{g} & $12.55\pm 0.08$ & $11.87 \pm 0.07$ & $11.58 \pm 0.05$ &\phn $9.72\pm 0.41$ & M7.0 & $23\pm4$ & $0.098_{-0.012}^{+0.022}$ & $1.9\pm 0.4$ & $9_{-2}^{+3}$ \\
2M\,1847B\tablenotemark{g} & $12.81\pm 0.20$ & $12.21 \pm 0.16$ & $11.74 \pm 0.11$ &\phn $9.88\pm 0.42$ & M7.5 && $0.094_{-0.013}^{+0.014}$      \\

\enddata
\tablenotetext{a}{M$_{K_s}$ = 7.65 + 2.13(J-$K_s$) with a rms $\sigma_{M_{Ks}}\,=\,0.33$ derived in Paper I (\S 4.2); relationship is valid for M6.5$<$SpT$<$L1.}
\tablenotetext{b}{Spectral type estimated by SpT\,=\,3.54M$_{K_s}$\,-\,27.20 with $\pm1.5$ spectral subclasses of error in these estimates as derived in paper I (\S 4.5). SpT = 10 is defined as an L0; valid for M6.5$<$SpT$<$L1.} 
\tablenotetext{c}{Distances based on M$_{K_s}$ as described in \S 4.2.} 
\tablenotetext{d}{Mass determination uses the models of \cite{cha00}.}
\tablenotetext{e}{Projected separations.}
\tablenotetext{f}{Periods include a 1.26 multiplication of the projected separations compensating for random inclinations and eccentricities \citep{fis92}.}
\tablenotetext{g}{This paper; otherwise, Paper I.}
\end{deluxetable}

\clearpage
\begin{deluxetable}{llllllc}
\tabletypesize{\scriptsize}
\tablecaption{All Known Resolved VLM Binaries\tablenotemark{a} \label{tbl-4}}
\tablewidth{0pt}
\tablehead{
\colhead{Name} &
\colhead{Sep.\tablenotemark{b}} &
\colhead{Est.} &
\colhead{Est. $M_A$}&
\colhead{Est. $M_B$}&
\colhead{Est. Period\tablenotemark{c}} &
\colhead{Ref.\tablenotemark{i} }\\ 
\colhead{} &
\colhead{AU} &
\colhead{$SpT_A/SpT_B$} &
\colhead{$M_{\sun}$} &
\colhead{$M_{\sun}$} &
\colhead{yr} &
\colhead{} 
}  
\startdata
PPL\,15\tablenotemark{d}   &  0.03 & M7/M8 & 0.07 & 0.06 & 5.8 days & 1 \\
Gl\,569B\tablenotemark{e}  &  1.0  & M8.5/M9.0 & 0.063 & 0.06 & 3 & 2,3 \\
SDSS\,2335583-001304 & 1.1? & L1?/L4? & 0.079 & 0.074 & 3 & 4 \\
2MASSI\,J1112256+354813 & 1.5 & L4/L6 &0.073 & 0.070 & 5 & 4 \\
2MASSI\,J1534498-295227 & 1.8 & T5.5/T5.5 & 0.05 & 0.05 & 8 & 5\\  
2MASSI\,J1847034+552243&  1.9 & M7/M7.5 & 0.098 & 0.094 & 9  & This paper\\
2MASSW\,J0856479+223518 & 2.0 & L5?/L8? & 0.071 & 0.064 & 8 & 4 \\
DENIS-P\,J185950.9-370632 & 2.0 & L0/L3 & 0.084 & 0.076 & 7 & 4 \\
HD\,130948B\tablenotemark{d} & 2.4 & L2/L2 & 0.07 & 0.06 & 10& 6 \\
$\epsilon$\,Indi B & 2.6 & T1/T6 & 0.042 & 0.027 & 16 & 16 \\
2MASSW\,J0746425+200032   &  2.7  & L0/L1.5 & 0.085 & 0.066 & 11 & 7, 17,20  \\
2MASSW\,J1047127+402644   &  2.7  & M8/L0 & 0.092 & 0.084 & 11 & 8, 17\\
DENIS-P\,J035726.9-441730 & 2.8 & L2/L4 & 0.078 & 0.074 & 12& 4,13 \\
2MASSW\,J12255432-2739466 &  3.2  & T6/T8 & 0.033 & 0.024 & 23  & 5 \\
2MASSW\,J0920122+351742   &  3.2  & L6.5/L7 & 0.068 & 0.068 & 16  & 7 \\
LHS\,1070B\tablenotemark{f} & 3.4 & M8.5/M9.0 & 0.070 & 0.068 & 16 & 18 \\
LP\,415-20 & 3.5 & M7/M9.5 & 0.095 & 0.079 & 15 & 9, Paper I \\
2MASSW\,J1728114+394859   &  3.7  & L7/L8 & 0.069 & 0.066 & 19  & 4,13\\
LHS\,2397a & 3.9 & M8/L7.5 & 0.090 & 0.068 & 22 & 10,17\\
2MASSW\,J1426316+155701   &  3.9  & M8.5/L1 & 0.088 & 0.076 & 19 & 17\\
2MASSW\,J2140293+162518   &  3.9 & M9/L2 & 0.092 & 0.078 & 22 & 17 \\
2MASSW\,J2206228-204705   &  4.4  & M8/M8 & 0.092 & 0.092 & 22 & 17\\ 
2MASSs\,J0850359+105716   &  4.4  & L6/L8 & 0.05 & 0.04 & 30 & 7 \\
2MASSW\,J1750129+442404   &  4.8  & M7.5/L0 & 0.095 & 0.084 & 25 & 9, Paper I \\
DENIS-P\,J1228.2-1547 & 4.9 & L5/L5 & 0.05 & 0.05 & 34 &  11 \\
2MASSW\,J1600054+170832   &  5.0  & L1/L3 & 0.078 & 0.075 & 29  & 4,13\\
2MASSW\,J1239272+551537   &  5.1  & L5/L5 & 0.071 & 0.071 & 31  & 4,13\\
2MASSI\,J0429184-312356 &  6.0 & M7.5/L1 & 0.094 & 0.079 & 50  & This paper\\
IPMBD\,29 & 7.2 & L1/L4\tablenotemark{g}  & 0.045 & 0.038 & 68 & 14 \\
2MASSW\,J1146345+223053	& 7.6 & L3/L4 & 0.055 & 0.055 & 63 & 12\\
2MASSW\,J1311391+803222  & 7.7 & M8.5/M9 & 0.089 & 0.087 & 51 & 17\\
CFHT-Pl-12 & 7.8 & M8/L4\tablenotemark{g} & 0.054 & 0.038 & 76 & 14 \\
2MASSW\,J1127534+741107  & 8.3 & M8/M9 & 0.092 & 0.087 & 57 & 17\\
LP\,475-855 & 8.3 & M7.5/M9.5 & 0.091 & 0.080 & 58 & 9, Paper I \\ 
DENIS-P\,J0205.4-1159 & 9.2 & L7/L7 & 0.07 & 0.07 & 75 & 15\\
2MASSW\,J2101349+175611	& 9.6 & L7/L8 & 0.068 & 0.065 & 82 &4,13\\
2MASSW\,J2147436+143131	&10.4 & L0/L2 & 0.084 & 0.078 & 83 &4,13\\
2MASSW\,J1449378+235537	&11.7 & L0/L3 & 0.084 & 0.075 &100 &4,13\\
IPMBD\,25 & 11.8 & M7/L4\tablenotemark{g} & 0.063 & 0.039 & 126 & 14 \\
DENIS-P\,J144137.3-094559 &13.5 & L1/L1 & 0.079& 0.079&124 &4,13\\
2MASSW\,J2331016-040618 \tablenotemark{e}  & 15.0 & M8.5/L7 & 0.093 & 0.067 & 159 & 17 \\
CFHT-Pl-18 & 34.5 & M8/M8 & 0.09 & 0.09 & 641 & 4,19 \\
2MASSW\,J1207334-393254 \tablenotemark{h} & 55 & M8/L5-L9.5 & 0.02 & 0.005 & 2600 & 22 \\
2MASSW\,J11011926-7732383 & 240 & M7/M8 & 0.05 & 0.025 & 13600 & 21 \\

\enddata
\tablenotetext{a}{We define VLM binaries as systems whose total mass is $<0.19\,M_{\sun}$. Very young evolving systems like GG TauBaBb \citep{whi01} are not included, nor are overluminous systems that are not resolved into binaries.}
\tablenotetext{b}{Projected separation except for the systems where semimajor axes have been measured.}
\tablenotetext{c}{This ``period'' is simply an estimate assuming a face-on circular orbit except for the systems where P has been measured.}
\tablenotetext{d}{PPL 15 is a spectroscopic binary (\cite{bas99} and
is not resolved).}
\tablenotetext{e}{These tight binaries have a widely separated more massive primary.}
\tablenotetext{f}{Part of a triple system.}
\tablenotetext{g}{Spectral type estimated from the given masses using Pleiades age, effective temperatures from \cite{bur01}, and spectral type-temperature relation for ultracool dwarfs from \cite{dah02}.}
\tablenotetext{h}{Parallaxes and common proper motions have yet to be confirmed}
\tablenotetext{i}{REFERENCES--(1)\cite{bas99}; (2) \cite{ken01}; (3)
\cite{lan01}; (4) \cite{bou03}; (5) \cite{bur03}; (6) \cite{pot02};
(7) \cite{rei01}; (8) \cite{rei02}; (9) \cite{sie03}; (10)
\cite{fre02}; (11) \cite{mar99}; (12) \cite{kor99}; (13) \cite{giz03};
(14) \cite{mar03}; (15) \cite{del97}; (16) \cite{mca03}; (17)
\cite{clo04}; (18) \cite{lei01}; (19) \cite{mar00}; (20) \cite{bou04};
(21) \cite{luh04}; (22) \cite{cha04}}
\end{deluxetable}

\clearpage

\clearpage

\clearpage

\clearpage

\clearpage

\clearpage

\clearpage

\clearpage

\end{document}